
\input harvmac
\noblackbox
\def\lae{\raise-.5ex\vbox{\hbox{$\; <\;$}\vskip-2.9ex\hbox{$\; \sim\;$}}}
\def\gae{\raise-.5ex\vbox{\hbox{$\; >\;$}\vskip-2.9ex\hbox{$\; \sim\;$}}}
\def\slash#1{\raise.15ex\hbox{/}\kern-.57em #1}

\def\ie{{\it i.e.}}

\def\two#1{\raise1.35ex\hbox{$\leftrightarrow$}\kern-.88em#1}
\def\lefta#1{\raise1.35ex\hbox{$\leftarrow$}\kern-.61em#1}
\def\righta#1{\raise1.35ex\hbox{$\rightarrow$}\kern-.61em#1}
\def\Dslash{\raise.15ex\hbox{/}\kern-.77em D}
\def\np#1#2#3{Nucl. Phys. {\bf #1} (#2) #3}
\def\pl#1#2#3{Phys. Lett. {\bf #1} (#2) #3}

\def\prd#1#2#3{Phys. Rev. D {\bf #1} (#2) #3}
\def\vsl{\raise.15ex\hbox{/}\kern-.57em v}
\def\ie{{\it i.e.}}

\def\tr{{\rm tr}}

\def\Boxmark#1#2#3{\global\setbox0=\hbox{\lower#1em \vbox{\hrule height#2em
     \hbox{\vrule width#2em height#3em \kern#3em \vrule width#2em}%
     \hrule height#2em}}%
     \dimen0=#2em \advance\dimen0 by#2em \advance\dimen0 by#3em
     \wd0=\dimen0 \ht0=\dimen0 \dp0=0pt
     \mkern1.5mu \box0 \mkern1.5mu }

\Title{\vbox{\baselineskip12pt\hbox{BUHEP-92-35}\hbox{HUTP-92/A058}
\hbox{hep-ph/9210276}}}
{Critical Constraints on Chiral Hierarchies}

\centerline{R. Sekhar Chivukula$^{a,1}$, Mitchell Golden$^{b,2}$, and
Elizabeth H. Simmons$^{b,3}$}
\footnote{}{$^a$Boston University, Department of Physics, 590 Commonwealth
Avenue, Boston, MA 02215}
\footnote{}{$^b$Lyman Laboratory of Physics,
Harvard University, Cambridge, MA 02138}
\footnote{}{$^1$sekhar@weyl.bu.edu}
\footnote{}{$^2$golden@physics.harvard.edu}
\footnote{}{$^3$simmons@physics.harvard.edu}
\vskip .4in
\centerline{\bf ABSTRACT}

We consider the constraints that critical dynamics places on models with a top
quark condensate or strong extended technicolor (ETC).  These models require
that chiral-symmetry-breaking dynamics at a high energy scale plays a
significant role in electroweak symmetry breaking.  In order for there to be a
large hierarchy between the scale of the high energy dynamics and the weak
scale, the high energy theory must have a second order chiral phase
transition. If the transition is second order, then close to the transition
the theory may be described in terms of a low-energy effective Lagrangian with
composite ``Higgs'' scalars.  However, scalar theories in which there are more
than one $\Phi^4$ coupling can have a {\it first order} phase transition
instead, due to the Coleman-Weinberg instability.  Therefore, top-condensate
or strong ETC theories in which the composite scalars have more than one
$\Phi^4$ coupling cannot always support a large hierarchy.  In particular, if
the Nambu--Jona-Lasinio model solved in the large-$N_c$ limit is a good
approximation to the high-energy dynamics, then these models will not produce
acceptable electroweak symmetry breaking.

\Date{10/92}

\newsec{Introduction}

Much recent work has focused on top quark condensate (and related) models
\ref\topmode{Y. Nambu, Enrico Fermi Institute Preprint EFI 88-39\semi V. A.
Miransky, M. Tanabashi, and K. Yamawaki, \pl{B221}{1989}{177} and Mod. Phys.
Lett. {\bf A4} (1989) 1043.}\nref\bardeen{W. A. Bardeen, C. T. Hill, and M.
Lindner, \prd{41}{1990}{1647}.}\nref\fourgen{C.~T.~Hill, M.~Luty, and
E.~A.~Paschos, \prd{43}{1991}{3011} \semi T.~Elliot and S.~F.~King,
\pl{B283}{1992}{371}.}\nref\brokentc{C.~T.~Hill, D.~C.~Kennedy, T.~Onogi, and
H-L.~Yu, Fermilab preprint FERMI-PUB-92/218-T.}--\ref\models{C.~T.~Hill,
\pl{B266}{1991}{419}\semi S.~Martin, \prd{45}{1992}{4283} and
\prd{46}{1992}{2197}\semi N.~Evans, S.~King, and D.~Ross, Southampton
University preprint SHEP-91-92-11.} as well as models with strong extended
technicolor interactions \ref\setc{T.~Appelquist, T.~Takeuchi, M.~Einhorn, and
L.~C.~R.~Wijewardhana, \pl{B220}{1989}{223}\semi T.~Takeuchi,
\prd{40}{1989}{2697}\semi V.~A.~Miransky and K.~Yamawaki, Mod. Phys. Lett.
{\bf A4} (1989) 129.}.  In these theories chiral symmetry breaking driven by
dynamics at a high scale ($\Lambda \gg$ 1 TeV) plays a significant role in
electroweak symmetry breaking.  Typically, the high-energy dynamics is
assumed to be a broken gauge theory -- either extended technicolor (ETC)
dynamics in strong ETC models or the dynamics of some grand unified theory in
top-condensate models.  The high-energy dynamics is usually modeled by a local
Nambu--Jona-Lasinio (NJL) four-fermion interaction \ref\nambu{Y. Nambu and G.
Jona-Lasinio, Phys. Rev.  {\bf 122} (1961) 345.} that is attractive in the
chiral-symmetry-breaking channel.  When the strength of the four-fermion
interaction is tuned close to the critical value for chiral symmetry breaking,
it would appear possible for the high-energy dynamics to play a role in
electroweak symmetry breaking without driving the electroweak scale to be of
order $\Lambda$.

The argument that high energy dynamics can play a role in electroweak symmetry
breaking is independent of the NJL approximation \ref\cohen{R.~S.~Chivukula,
A.~G.~Cohen, and K.~Lane, \np{B343}{1990}{554}.}: If the coupling constants of
the high energy theory are small, only strong low-energy dynamics (such as
technicolor) can contribute to electroweak symmetry breaking. On the other
hand, if the coupling constants of the high-energy theory are large and the
interactions are attractive in the appropriate channels, chiral symmetry will
be broken by the high-energy interactions and the scale of electroweak
symmetry breaking will be of order $\Lambda$.  If the transition between these
two extremes is continuous, \ie\ if the chiral symmetry breaking phase
transition is {\it second order} in the high-energy couplings, then it is
possible to adjust the high energy parameters so that the dynamics at scale
$\Lambda$ can contribute to electroweak symmetry breaking.  Moreover, if the
transition is second order, then close to the transition the theory may be
described in terms of a low-energy effective Lagrangian with composite
``Higgs'' scalars -- the Ginsburg-Landau theory of the chiral phase
transition.

It is crucial that the transition be second order in the high energy
couplings.  If the transition is first order, then as one adjusts the
high-energy couplings the scale of chiral symmetry breaking will jump
discontinuously from approximately zero at weak coupling to approximately
$\Lambda$ at strong coupling.  In general it will not be possible to maintain
a hierarchy between the scale of electroweak symmetry breaking and scale of
the high energy dynamics, $\Lambda$.

In this note we show that there are cases in which the transition cannot be
self-consistently second order.  A scalar theory in which there is more than
one $\Phi^4$ coupling can have a {\it first order} phase transition instead,
due to the Coleman-Weinberg instability \ref\cw{S. Coleman and E. Weinberg,
\prd{7} {1973} {1888}.}.  Therefore, top-condensate or strong ETC theories in
which the composite scalars have more than one $\Phi^4$ coupling cannot always
support a large hierarchy.

\newsec{$U(N_f) \times U(N_f)$ models}

For simplicity, we first consider a theory of $N_f$ left- and right-handed
fermions $\Psi$ with a chiral $U(N_f) \times U(N_f)$ symmetry.  As usual, we
assume that the high-energy dynamics is attractive in the $\bar{\Psi} \Psi$
channel.  Therefore, the order parameter $\Phi$ of chiral symmetry breaking
transforms as an $(\overline{N}_f, N_f)$ under the chiral symmetry. If it is
possible to arrange for a large hierarchy, then at energies below $\Lambda$ the
dynamics can be described in terms of a Ginsburg-Landau theory for the order
parameter $\Phi$ coupled to the fermions
\eqn\efflag{
\eqalign{
\CL = & \overline\Psi i\Dslash \Psi
+ {\pi y \over \sqrt{N_f}} \left(\overline\Psi_{L} \Phi \Psi_R  + h.c. \right)
+ \tr(\partial^\mu \Phi^\dagger \partial_\mu \Phi) -
 M^2 \tr(\Phi^\dagger \Phi) \cr
& - {\pi^2 \over 3} {\lambda_1 \over N^2_f} (\tr \Phi^\dagger \Phi)^2
 - {\pi^2 \over 3} {\lambda_2 \over N_f} \tr (\Phi^\dagger \Phi)^2
+ O\left({{\Phi^\dagger \Phi}\over \Lambda^2}, {\partial^2\over
\Lambda^2}\right)~.
}
}
The quantities $y$, $M^2$, $\lambda_1$ and $\lambda_2$ are functions
of the couplings of the fundamental high-energy theory. This effective
Lagrangian can be considered the theory of a composite $U(N_f) \times
U(N_f)$ ``Higgs'' boson $\Phi$.

At tree level, if the high-energy couplings can be chosen so that $M^2 \ll
\Lambda^2$, then it is possible to establish a large hierarchy.  This
prediction can be affected by quantum corrections: as shown by Coleman and
Weinberg \cw, if $M^2$ is adjusted to be close to zero, then quantum
corrections can destabilize the minimum at $\Phi = 0$. More precisely, if one
computes the renormalization-group-improved effective potential and requires
that the second derivative of the potential at $\Phi=0$ be small, one finds
that the potential is minimized far away from the origin.  Consequently, if
one adjusts the couplings in the high-energy theory so that $M^2$ goes through
zero, one finds that the location of the effective potential's absolute
minimum jumps discontinuously from $\Phi=0$ to some large nonzero value of
$\Phi$.  In other words, the transition which at tree level was second order
is driven first order by quantum fluctuations\foot{The stability of the
$U(N_f)\times U(N_f)$ linear sigma-model, without fermions, was considered in
\ref\paterson{A. J. Paterson, Nucl.  Phys.  {\bf B190} [FS3] (1981) 188.}.}.

Yamagishi \ref\yama{H. Yamagishi, \prd{23}{1981}{1880}.} has shown that the
condition that the effective potential be minimized away from the origin can be
stated purely in terms of the couplings $\lambda_1(\mu)$ and $\lambda_2(\mu)$,
by following their flows from $\mu \approx \Lambda$ as the scale $\mu$ is
decreased.  We apply the results of \yama\ to the Lagrangian \efflag.  The
effective potential is minimized away from the origin if the couplings
cross the line
\eqn\conditiona{
4(\lambda_1 + \lambda_2) + \beta_1 + \beta_2  = 0
}
in a region where $\lambda_2>0$, $\lambda_1 + \lambda_2 < 0$ and
\eqn\conditionb{
4(\beta_1 + \beta_2) + \sum_{i,j=1,2} \beta_i  {\partial
\beta_j\over \partial \lambda_i} > 0~.
}
Here $\beta_1$ and $\beta_2$ are the beta functions for the couplings
$\lambda_1$ and $\lambda_2$, respectively.  We will refer to the line
\conditiona\ as the ``stability line''.

If the couplings never cross the stability line, quantum corrections do not
drive the transition first order and the high-energy theory may
self-consistently have a second order transition.  However, if the couplings
do cross the stability line, the low-energy effective theory has a first-order
transition and therefore the high-energy theory cannot self-consistently have
a second order transition \ref\amit{See, for example, D.~J.~Amit, {\it Field
Theory, the Renormalization Group, and Critical Phenomena}, 2nd ed., World
Scientific, Singapore, 1984.}.

In practice, of course, one can only compute the beta functions in
perturbation theory.  At one-loop the beta functions are\foot{The
contributions from $\lambda_1$ and $\lambda_2$ differ from those given in
\ref\pisarski{R.~D.~Pisarski and D.~L.~Stein, Phys. Rev. B. {\bf 23} (1981)
3549 and J. Phys. A. {\bf 14} (1981) 3341.} and \paterson\ by a factor of 1/4.
We note that in \paterson, the complex scalar field is incorrectly normalized
and this explains the discrepancy.}
\eqn\bfunctioni{\beta_1 =  {1\over 12} \left( 1 + {4\over
N^2_f}\right) \lambda^2_1 + {1\over 3} \lambda_1 \lambda_2 + {1\over 4}
\lambda^2_2
+{y^2 N_c \over 4 N_f} \lambda_1,}
and
\eqn\bfunctionii{\beta_2 = {1\over 2 N^2_f} \lambda_1 \lambda_2 + {1\over
6}\lambda^2_2
+{y^2 N_c \over 4 N_f} \lambda_2 - {3 N_c \over 8 N_f} y^4.}
Here $N_c$ is the number of colors or technicolors of fermions $\Psi$.  Note
that, if $y$ is constant, the one-loop $\beta$-functions for the quantities
$\lambda_1/y^2$ and $\lambda_2/y^2$ are independent of $y$.  The
$\beta$-functions \bfunctioni\ and \bfunctionii\ have a fixed point which is
the analog of the fixed point for the Higgs self-coupling noted in \bardeen.

In these theories, typically the Yukawa coupling is drawn quickly to a
low-energy ``fixed point'' \ref\pross{B. Pendleton and G.  Ross,
\pl{98B}{1981}{291}.} \ref\hillfix{C. Hill, \prd{24}{1981}{691} \semi M.
Fischler and C.  Hill, \np{B193}{1981}{53}.}, where its value runs very slowly
due to the running of a relatively weak gauge coupling (color or technicolor).
For the purposes of illustration, therefore, we ignore the running of the
Yukawa coupling $y$.

In \fig\figa{The trajectories for the quantities $\lambda_1/y^2$ and
$\lambda_2/y^2$ in the $U(N_f) \times U(N_f)$ model.  The arrows indicate the
behavior as one scales toward the infrared.  Here we have taken $N_f= 2$ and
$N_c = 3$.  Because of the form of equations \bfunctioni\ and \bfunctionii,
this plot is independent of $y$.  The ``stability line'' is shown in dashes.
Note that the curves that start at large $\lambda_2$ and small $\lambda_1$
cross the stability line twice, and thus have a Coleman-Weinberg instability.}
we plot the renormalization group trajectories of $\lambda_1/y^2$ and
$\lambda_2/y^2$ for the model with $N_f=2$ and $N_c=3$.  These figures show
that the couplings $\lambda_1$ and $\lambda_2$ run toward the fixed point of
\bfunctioni\ and \bfunctionii\ discussed above.  If $\lambda_2>\lambda_1$ and
if both are sufficiently strong at $\mu=\Lambda$, the couplings run in such a
way as to intersect the stability line.  In fact, these trajectories intersect
the line {\it twice}. One can check that, as one scales to the infrared
(toward the fixed point), condition \conditionb\ is satisfied only at the {\it
first} intersection and this intersection corresponds to the minimum of
the effective potential.  We have numerically checked that the picture does
not qualitatively change with a running Yukawa coupling or for different
values of $N_f$ and $N_c$.

In the cases where the two $\lambda$'s start at reasonably large values, they
run quickly and intersect the stability line after a small change in $\mu$.
At one loop, the value of $\Phi$ at the minimum of the potential is equal to
the value of $\mu$ at which the stability line is crossed.  Therefore, if the
couplings cross the stability line quickly, then $\langle \Phi \rangle$ is of
order $\Lambda$ and there can be no large hierarchy.

Of crucial importance, then, is what values the couplings $\lambda_1(\mu)$ and
$\lambda_2(\mu)$ take when $\mu = \Lambda$.  This is a non-perturbative
problem.  In the NJL model one may show \bardeen\ that to leading order in
$1/N_c$, $\lambda_1(\mu) \rightarrow 0$ and $\lambda_2(\mu) \rightarrow
\infty$ as $\mu \rightarrow \Lambda$.  This boundary condition puts the
$U(N_f) \times U(N_f)$ model in the region which flows rapidly toward the
stability line and therefore suggests that it is not possible to obtain a
large hierarchy\foot{These predictions will be modified in a generalized NJL
model \ref\shen{A.~Hasenfratz, P.~Hasenfratz, K.~Jansen, J.~Kuti and Y.~Shen,
\np{B365}{1991}{79}.}. However, even in generalized models $\lambda_1(\mu)
\rightarrow 0$ as $\mu \rightarrow \Lambda$ to leading order in $1/N_c$.
Therefore, we expect that the transition will still be first order if
$\lambda_2(\Lambda)$ is not small.}.

One may be concerned that we are investigating the Coleman-Weinberg phenomenon
in perturbation theory, but have been forced to consider potentially large
values of the couplings $\lambda$. However, since the phenomenon depends only
on the {\it qualitative} features of the renormalization group flows, we do
not expect that higher order effects will qualitatively change the
conclusions. This issue may be tested by simulating the model \efflag\
nonperturbatively using lattice techniques. While this has not been done in
four dimensions, numerical simulations in three dimensions without fermions
(where the lowest-order renormalization group analysis also predicts a first
order transition \ref\wilczek{R.~D.~Pisarski and F.~Wilczek,
\prd{29}{1984}{338}.}) confirm that the transition is first order
\ref\gauseterer{H. Gausterer and S. Sanielevici, \pl{209B}{1988}{533}.}.

The point is that it is not sufficient to adjust the couplings of the
high-energy theory so that the second derivatives of the scalar potential at
the origin are small. One will also have to adjust the theory so that, at $\mu
\approx \Lambda$, one is in a region of coupling constant space which does not
quickly flow toward the stability line. In a spontaneously broken gauge theory
with a simple gauge group, however, having fixed the scale of symmetry
breaking one can only adjust {\it one} parameter: the value of the gauge
coupling at the symmetry breaking scale.  One cannot, therefore, simply assume
that a large hierarchy of scales is possible.  One must check that the
effective low-energy theory does not suffer from a Coleman-Weinberg
instability.  As we have seen the large-$N_c$ limit of the high-energy NJL
model places the $U(N_f) \times U(N_f)$ low-energy model in a region which has
this instability.

\newsec{Other models}

We now consider some other examples.  Consider first a generic theory without
fermions.  As before, we can introduce a field $\Phi$ to represent the order
parameter of chiral symmetry breaking.  If the symmetry of the high-energy
theory is such that the Ginsburg-Landau theory for $\Phi$ has more than one
coupling of dimension four, then, at least in the $\epsilon$-expansion, the
only fixed point is the infrared-unstable Gaussian fixed point.  One therefore
expects that the couplings generally flow toward the unstable region, \ie\
most trajectories are pushed away from the origin and flow toward large
negative values of the couplings.

Now consider the theory with fermions.  As we have seen, there will in general
be infrared-stable fixed-points. However, if the scalar self-couplings are
large compared to the Yukawa couplings, the coupling constant flows will (at
least initially) look the same as they did without fermions and should,
therefore, still cross the stability line.

Accordingly, in a model of composite scalars in which there is more than one
$\Phi^4$ coupling and in which the scalar self-interactions become strong at
the compositeness scale $\Lambda$, the chiral phase transition may not be
second order. Such a model will not always sustain a large hierarchy between
the compositeness-scale $\Lambda$ and the weak scale.

In top-condensate-inspired models with two composite ``Higgs'' bosons
\ref\luty{M.~A.~Luty, \prd{41}{1990}{2893}.} \ref\suzfro{M.~Suzuki,
\prd{41}{1990}{3457} \semi C.~D.~Froggatt, I.~G.~Knowles, and R.~G.~Moorhouse,
\pl{B249}{1990}{273}.}, for example, one has five $\Phi^4$ couplings and three
mass terms. It can be argued that one has enough freedom to adjust the three
mass terms to be close to zero, but for the reasons discussed above the theory
can still have a fluctuation-induced first-order phase transition.  Again, if
large-$N_c$ arguments apply, the model will not sustain a large
hierarchy\foot{The instability was noted in \luty, but its implications were
not discussed.}.

By contrast, the standard $O(4)$ model \topmode\ has only one quartic
coupling.  In this case, the ``stability line'' is a point, and it is at a
{\it lower} value of $\lambda$ than the fixed point.  Therefore, if, as in
\bardeen, the value of $\lambda(\Lambda)$ is large, then the trajectory hits
the fixed point without crossing the stability point and it may be possible to
sustain a large hierarchy.

Note that our results apply only in cases in which the scalar
self-interactions become strong at the compositeness scale.  In
composite-Higgs models in which {\it all} of the scalars are Goldstone Bosons
of some chiral symmetry breaking transition at a higher energy scale
\ref\kaplan{D.~B.~Kaplan and H.~Georgi, \pl{136B}{1984}{183}.}, the
nonderivative self-couplings of the scalars are related to small symmetry
breaking effects and can naturally be small at $\mu = \Lambda$.  Although the
transition may in principle be first order, it may take a very large change of
scale before the couplings cross the stability line since the couplings are
weak. In this case the hierarchy can be large.

\newsec{Conclusions}

In conclusion, theories of composite ``Higgs'' scalars may have a first order
chiral symmetry breaking phase transition if there is more than one $\Phi^4$
coupling and if the scalar self-interactions become strong at the
compositeness scale.  One must check that the theory does not suffer from the
Coleman-Weinberg instability.  In particular, in strong ETC models or
generalized top-condensate models with more than one $\Phi^4$ coupling in the
low-energy theory, one may not be able to adjust the high-energy theory to
obtain a large hierarchy between the scale of the high-energy dynamics and the
weak scale.  If the NJL model solved in the large-$N_c$ limit is a good
approximation to the high-energy dynamics, then these models will not produce
acceptable electroweak symmetry breaking.

\newsec{Acknowledgements}

We would like to thank Andrew Cohen, Mike Dugan, Marty Einhorn, Howard Georgi,
and Chris Hill for useful conversations and comments.  R.S.C. and E.H.S. thank
Robert Jaffe and Emil Mottola for organizing the {\it Sante Fe Workshop on
Hadrons and Physics Beyond the Standard Model} where some of this work was
completed.  R.S.C.  acknowledges the support of an Alfred P. Sloan Foundation
Fellowship, an NSF Presidential Young Investigator Award, and a DOE
Outstanding Junior Investigator Award.  R.S.C. and M.G. acknowledge
support from the Texas National Research Laboratory Commission under a
Superconducting Super Collider National Fellowship.  This work was supported
in part under NSF contracts PHY-90-57173 and PHY-87-14654 and DOE contracts
DE-AC02-89ER40509 and DE-FG02-91ER40676, and by funds from the Texas National
Research Laboratory Commission under grants RGFY92B6 and RGFY9206.

\listrefs
\listfigs
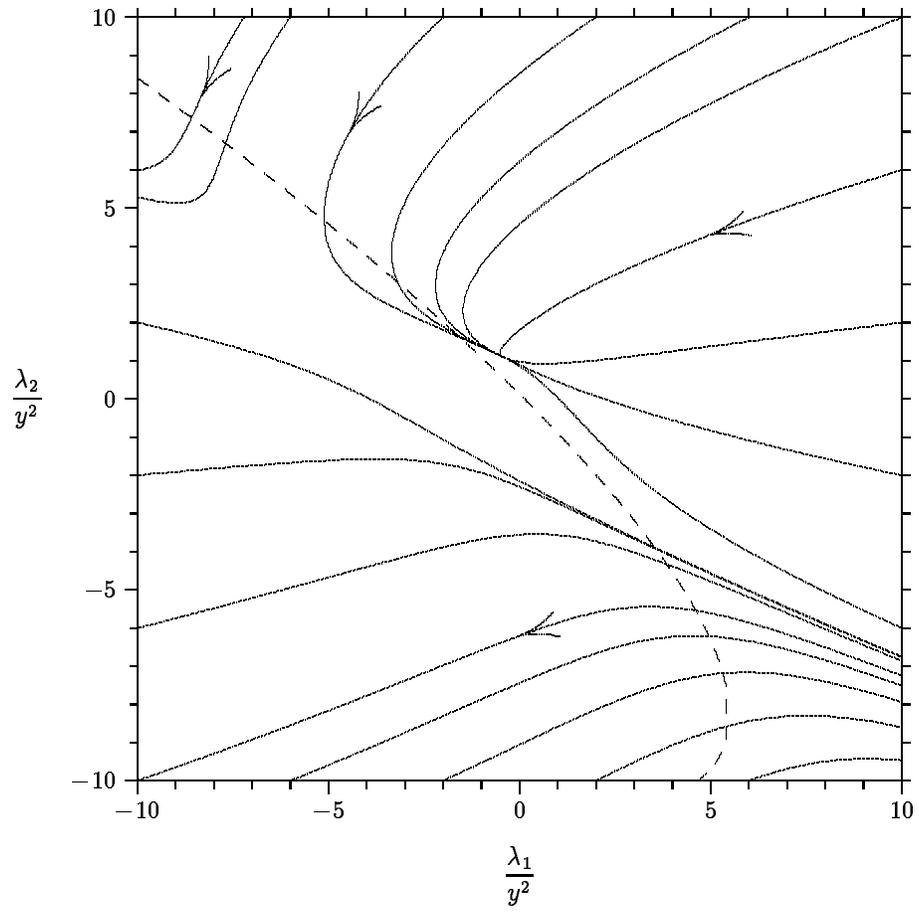
\bye